\documentclass[a4paper,11pt]{article}
\usepackage{pos}
\usepackage{braket}
\usepackage{xcolor}
\usepackage[normalem]{ulem}
\usepackage{enumitem}
\usepackage{hyperref}
\usepackage{natbib}
\usepackage{caption}
\usepackage{subcaption}

\title{$N_f = 1+2+1$ QCD+QED simulations with C$^\star$ boundary conditions}

\author[a]{Anian Altherr}
\author[b]{Lucius Bushnaq} 
\author[c]{Isabel Campos} 
\author[a]{Marco Catillo} 
\author[d]{Alessandro Cotellucci}
\author[d,e,f,g]{Madeleine Dale}
\author[b]{Patrick Fritzsch}
\author[a]{Roman Gruber}
\author*[d,h]{Jens L\"ucke}
\author[a]{Marina Krstić Marinković}
\author[d,h]{Agostino Patella}
\author[e,f]{Nazario Tantalo}
\author[a]{Paola Tavella}

\affiliation[a]{Institut f\"ur Theoretische Physik, ETH Z\"urich, Wolfgang-Pauli-Str. 27, 8093 Z\"urich, Switzerland}
\affiliation[b]{School of Mathematics, Trinity College Dublin, Dublin 2, Ireland}
\affiliation[c]{Instituto de F\'isica de Cantabria \& IFCA-CSIC, Avda. de Los Castros s/n, 39005 Santander, Spain}
\affiliation[d]{Humboldt Universit\"at zu Berlin, Institut f\"ur Physik \& IRIS Adlershof, \\Zum Grossen Windkanal 6, 12489 Berlin, Germany}
\affiliation[e]{Universit\`a di Roma Tor Vergata, Dipartimento di Fisica, \\Via della Ricerca Scientifica 1, 00133 Rome, Italy}
\affiliation[f]{INFN, Sezione di Tor Vergata, Via della Ricerca Scientifica 1, 00133 Rome, Italy}
\affiliation[g]{University of Cyprus, Department of Physics, 1 Panepistimiou Street, 2109 Aglantzia, Nicosia, Cyprus}
\affiliation[h]{DESY, Platanenallee 6, D-15738 Zeuthen, Germany}

\emailAdd{jens.luecke@hu-berlin.de}

\abstract{

\begin{center}
\large
\includegraphics[width=.11\textwidth]{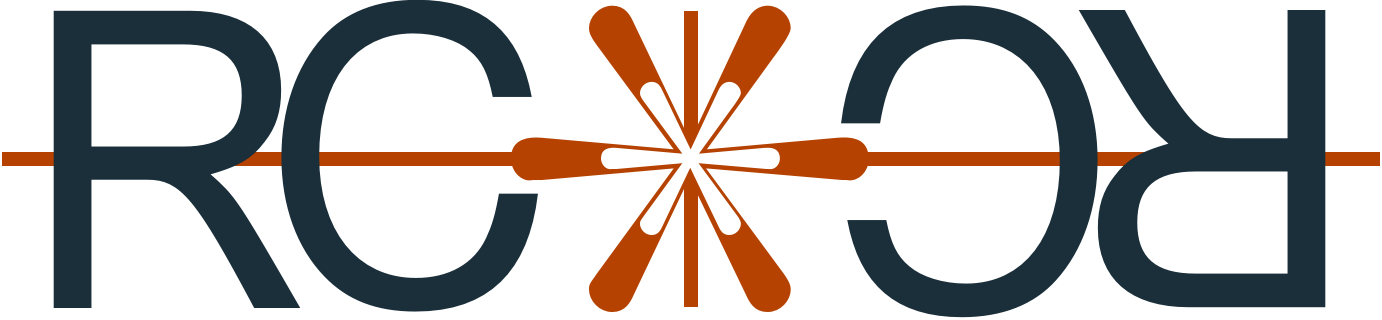} \hspace{1ex} collaboration
\end{center}

\bigskip
We give an update on the ongoing effort of the RC$^\star$ collaboration to generate fully dynamical QCD+QED configurations with C$^\star$ boundary conditions using the openQ$^\star$D code. The simulations are tuned to the U-symmetric point ($m_d=m_s$) with pions at $m_{\pi^\pm}\approx 400$ MeV. The splitting of the light mesons is used as one of three tuning observables and fixed to $m_{K^0} - m_{K^\pm} \approx 5$ MeV and $m_{K^0} - m_{K^\pm} \approx 25$ MeV on ensembles with renormalized electromagnetic coupling $\alpha_\mathrm{R} \approx \alpha_\mathrm{phys.}$ and $\alpha_\mathrm{R}\approx 5.5 \alpha_\mathrm{phys.}$ respectively. We will discuss some details concerning our tuning strategy and present the calculation of the meson and baryon masses. Finally, we will also present a cost analysis for our simulations. More technical details on finite-volume effects and the tuning can be found in A. Cotellucci's proceedings~\cite{RC:2022fly}.}

\FullConference{%
The 39th International Symposium on Lattice Field Theory,\\
8th-13th August, 2022,\\
Rheinische Friedrich-Wilhelms-Universität Bonn, Bonn, Germany
}

\begin{document}
\maketitle

%
\section{Introduction}
We present an update on a long-term research program aiming at calculating isospin-breaking and QED radiative corrections in hadronic quantities, with C$^\star$ boundary conditions~\cite{Kronfeld:1990qu,Kronfeld:1992ae,Wiese:1991ku,Polley:1993bn} and fully-dynamical QCD+QED simulations.
C$^\star$ boundary conditions allow for a local and gauge-invariant formulation of QED in finite volume and in the charged sector of the theory~\cite{Lucini:2015hfa,Patella:2017fgk,Hansen:2018zre}. In particular, three new ensembles were generated at the values of the fine-structure constant $\alpha_R \approx 0.04$ and $\alpha_R \approx 1/137$. The larger value of $\alpha_R$ was chosen to amplify QED corrections.

The open-source \texttt{openQ*D-1.1} code~\cite{openQxD-csic,Campos:2019kgw} was used to generate all gauge configurations presented in this work. This code has been developed by the RC$^\star$ collaboration. It is an extension of the \texttt{openQCD-1.6} code~\cite{openQCD} for QCD.

In this proceedings, we will give a very broad overview of the ensembles we have generated and some observables that were computed on them. Starting from our renormalization scheme (section \ref{sec:ensembles}) we will show some baryon masses computed at physical electromagnetic coupling (section \ref{sec:masses}). This is followed by a short overview of the cost of the fully dynamical QCD+QED simulations we have done so far (section \ref{sec:cost}). The section on the finite volume effects, mentioned in the talk can be found in Alessandro Cotellucci's proceedings \cite{AlessandroPOS}.

%
\section{Ensembles and Setup}
\label{sec:ensembles}
We have generated two $N_f=3+1$ QCD ensembles and five $N_f=1+2+1$ QCD+QED ensembles. For the SU(3) field the L\"uscher-Weisz action with $\beta=3.24$ was used, while for the U(1) field the Wilson action with $\alpha_0 = 0.05$ and $\alpha_0 = 1/137$ (for the QCD+QED ensembles), and $O(a)$-improved Wilson fermions. We employ C$^\star$ boundary conditions in all three spatial and periodic boundary conditions in the time direction for all our ensembles. We have verified that we are free from the problem of topological freezing in all our ensembles, which justifies the use of periodic boundary conditions in time. More information on the setup and choice of parameters can be found in \cite{RC:2021tah,Bushnaq:2022aam}.

\begin{figure}
    \centering
    \includegraphics[width=\textwidth]{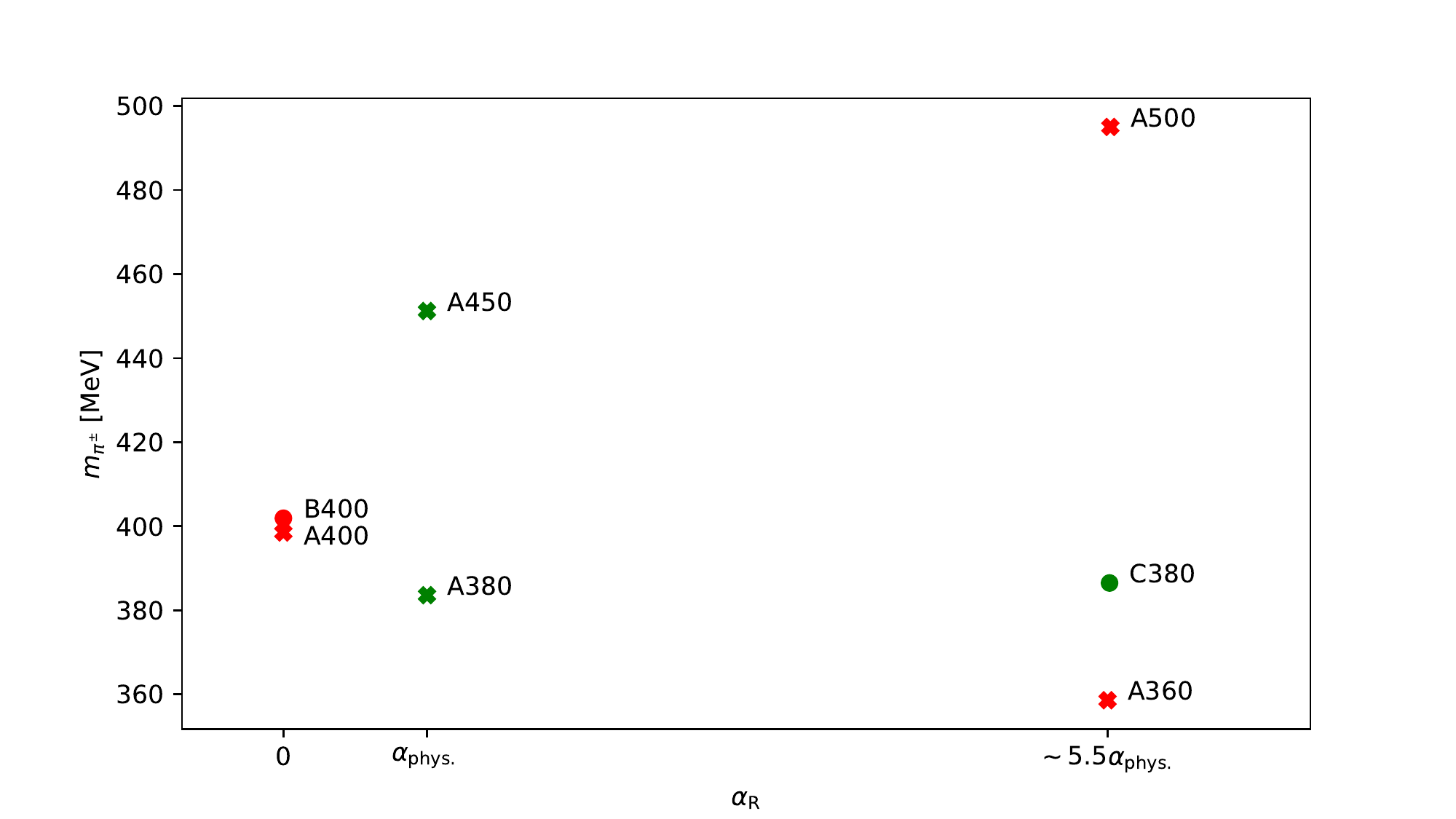}
    \caption{Overview of all ensembles generated so far. The first letter in the naming convention denotes the geometry of the ensemble. For example, A corresponds to a $64\times 32^3$ lattice, while C stands for a $96\times 48^3$ lattice. The number after the first letter stands for the mass of the lightest excitation in the system, i.e. the pion. Dots denote large volumes, while crosses denote the smaller $64\times 32^3$ lattices. In green are the new ensembles, while the red ones were already discussed in \cite{RC:2021tah}.}
    \label{fig:ensembles}
\end{figure}

\begin{figure}
    \centering
    \includegraphics[width=\textwidth]{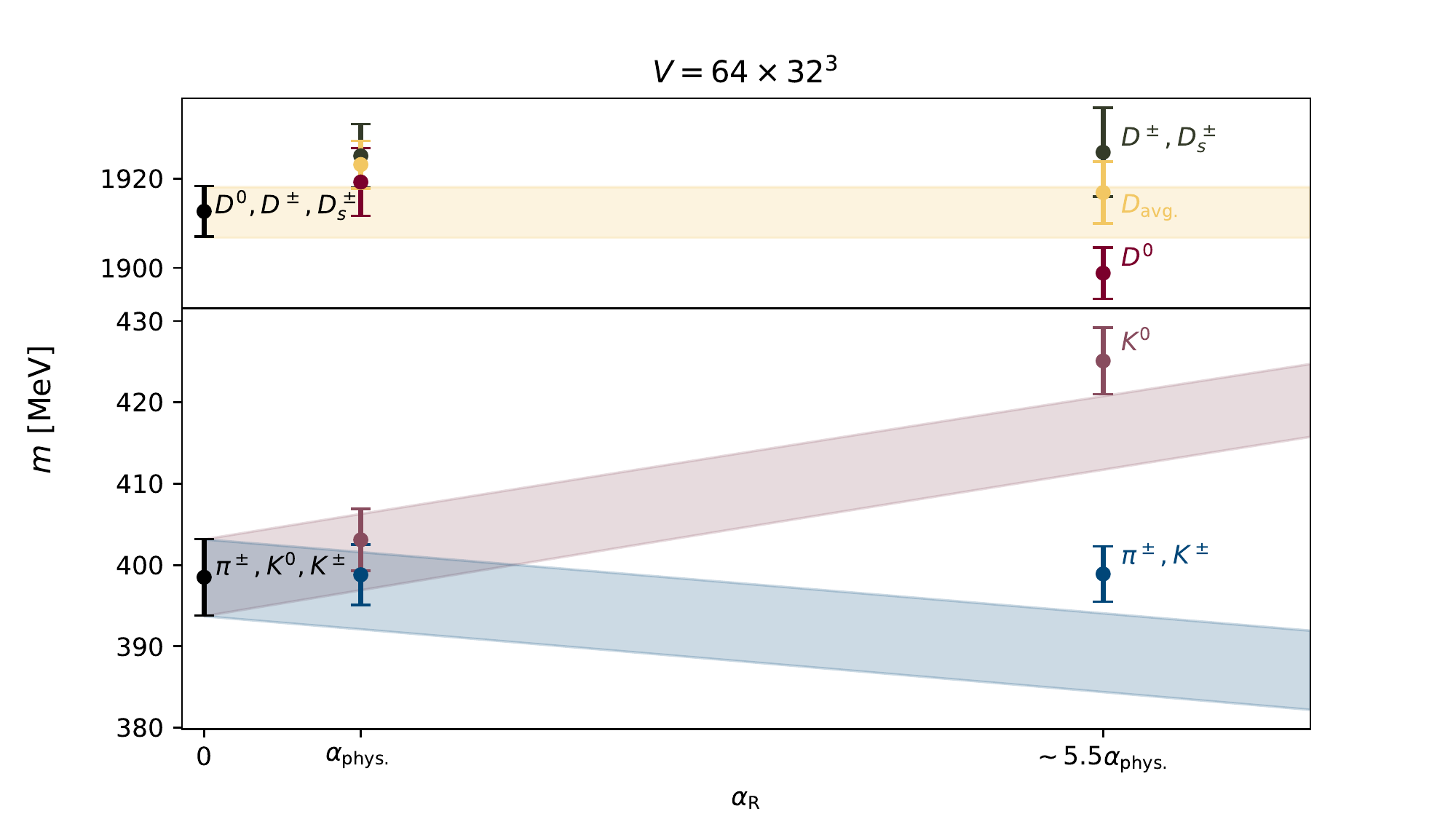}
    \caption{Illustration of the tuning procedure. On the left in black one can see the light meson masses in pure QCD at $\alpha_\mathrm{R}=0$, at the SU(3)-symmetric point (thus the degeneracy of the low-lying mesons). The yellow, red, and blue bands depict the lines of constant physics for the light (lower section) and heavy (upper section) mesons. On the QCD+QED ensemble at $\alpha_\mathrm{R}\approx 1/137$ one can see, that the light mesons are straight on the line of constant physics, while the heavy ones are slightly too heavy. The opposite situation is the case for the ensemble at the artificially large value of $\alpha_\mathrm{R}\approx 0.04$.}
    \label{fig:tuning}
\end{figure}

Compared to \cite{Bushnaq:2022aam}, the major updates are three new ensembles. There are two new ones at $\alpha_0 = 1/137$ on a $64 \times 32^3$ lattice and one at $\alpha_0 = 0.05$ on a $96\times 48^3$ lattice. An overview of all ensembles can be found in figure \ref{fig:ensembles}.
QCD+QED with four fermion flavors is parameterized by six inputs. The parameters of our ensembles are fixed via the (unphysical) scheme
\begin{subequations}
    \begin{align}
    \sqrt{8t_0} &= 0.415 \, \mathrm{fm} 
    &&&
    \alpha_\mathrm{R} &\in \{ 0, 1/137, 0.05\}
    \\
    \phi_0 &= 8t_0 \left( M_{\pi^\pm}^2 - M_{K^\pm}^2 \right) \overset{!}{=} 0
    &&&
    \phi_1 &= 8t_0 \left( M_{K^0}^2 + M_{\pi^\pm}^2 + M_{K^\pm}^2 \right) \overset{!}{=} \phi_1^\mathrm{phys.}
    \\
    \phi_2 &= 8t_0 \left( M_{K^0}^2 - M_{K^\pm}^2 \right)/\alpha_\mathrm{R} \overset{!}{=} \phi_2^\mathrm{phys.}
    &&&
    \phi_3 &= \sqrt{8t_0} \left( M_{D^0} + M_{D^\pm} + M_{D_s^\pm} \right) \overset{!}{=} \phi_3^\mathrm{phys.} \ .
    \end{align}
    \label{eq:scheme}
\end{subequations}
Taking the meson masses from the PDG \cite{ParticleDataGroup:2020ssz}, the value of $\sqrt{8t_0}$ from \cite{Bruno:2016plf} and the Thompson limit of the electromagnetic coupling $\alpha_\mathrm{R} = 1/137$ one gets 
\begin{align}
    \phi_0^\mathrm{phys.} = 0.992 \ ,
    &&&
    \phi_1^\mathrm{phys.} = 2.26 \ ,
    &&&
    \phi_2^\mathrm{phys.} = 2.36 \ ,
    &&&
    \phi_3^\mathrm{phys.} = 12.0 \ .
\end{align}
The scheme defined in eq. \eqref{eq:scheme} is unphysical in multiple ways. First of all, the overall scale $t_0$ cannot be measured directly in an experiment. Furthermore, choosing $\phi_0 = 0$ implies that the down and strange quark need to be degenerate. At some point in the future, the goal is to replace the scale $t_0$ by the mass of the omega baryon $M_{\Omega^-}$ and split the down and strange quark to be able to do simulations closer to the physical point. A plot of the chosen lines of constant physics can be found in figure \ref{fig:tuning}.

%
\section{Hadron masses}%
\label{sec:masses}
Both the meson and baryon masses are extracted in a gauge-invariant way by dressing the bare quark fields with Dirac factors \cite{Hansen:2018zre}. For the mesons, this is sufficient to build interpolating operators that have a good overlap with the meson states on our ensembles. For the baryons, the procedure is slightly more involved. Baryon operators are constructed from smeared, U(1)-gauge-invariant fermion fields and smeared gauge links. Gau{\ss}ian smearing is used for gauge invariant, dressed quark operator $\Psi$
\begin{align}
    \Psi_{\text{(s)}} 
    &= 
    ( 1 + \omega H[U_s])^n \Psi 
    \ .
\end{align}
Here $H[U_s]$ is the spatial hopping operator depending on smeared gauge field $U_s$. The gauge smearing is done with a modified gradient flow defined by the flow equation
\begin{align}
    \partial_s U_s(x,k) 
    &= 
    \partial_{x,k} \sum_{i \neq j} \mathrm{tr} P_{s,ij}(x) 
    &
    &\mathrm{with}
    &
    U_0 
    &= 
    U
    \ .
\end{align}
The spatial plaquettes in the point $x$, built from U(1)-gauge-links at positive flow-time $s$ are denoted by $P_{s,ij}(x)$.
The flowtime $s$ and parameters $n$ and $\omega$ are optimized using the GEVP. The modified version of the flow is used since it preserves locality in time. The interpolating operator for a spin-1/2 baryon is then given by
\begin{align}
    B(x_0) = \sum_{\vec{x}} \sum_{\substack{abc\\fgh}}
   \epsilon_{abc} F_{fgh} 
   \left\{ \Psi_{\text{(s)}fa} \Psi_{\text{(s)}gb}^t C \gamma_5 \Psi_{\text{(s)}hc}(x_0,\vec{x})
   - C \bar{\Psi}^t_{\text{(s)}fa} \bar{\Psi}_{\text{(s)}gb} C \gamma_5 \bar{\Psi}^t_{\text{(s)}hc}(x_0,\vec{x}) \right\} 
   \ .
   \label{eq:baryon:intop:B}
\end{align}
Here, $a,b,c$ are color indices and $f,g,h$ are flavor indices (spin indices
are implicit or contracted). The matrix $C$ is the charge conjugation operator acting on the spin indices.
Different baryons are obtained by choosing particular tensors $F_{fgh}$,
according to the table~\ref{tab:baryons1/2}.
\begin{table}
   \begin{center}
      \begin{tabular}{cc}
         $(1/2)^+$ Baryon & Non-zero components of $F_{fgh}$ \\
         \hline
         p & $F_{uud} = 1$ \\
         n & $F_{ddu} = 1$ \\
         $\Lambda_0$ & $F_{sud} = 2$, $F_{dus} = 1$, $F_{uds} = -1$ \\
         $\Sigma^+$ & $F_{uus} = 1$ \\
         $\Sigma^-$ & $F_{dds} = 1$ \\
         $\Xi_0$ & $F_{ssu} = 1$ \\
         $\Xi^-$ & $F_{ssd} = 1$ \\
         \hline
      \end{tabular}
   \end{center}
   \caption{%
   Flavor tensor $F_{fgh}$ defining the interpolating operators for spin-$1/2$
   baryons via eq.~\eqref{eq:baryon:intop:B}. The flavor indices can take
   values $u$, $d$, $s$, and $c$.
   }
   \label{tab:baryons1/2}
\end{table}
Projecting the two-point function onto the positive parity component, the $1/2^+$ correlation functions at zero momentum are
\begin{align}
   C(x_0) = \langle B^t(0) C \frac{1 + \gamma_0}{2} B(x_0) \rangle
   \ .
\end{align}
Working out the quark contractions, it turns out, that the correlation function decomposes into a single-quark-line-connected $C_1(x_0)$ and a three-quark-line-connected part $C_3(x_0)$. The contribution from $C_1$ is exponentially suppressed with the volume: With C$^\star$ boundary conditions, one has quark-quark and antiquark-antiquark contractions, because the quark and antiquark are related via the boundary conditions. Such flavor-violating artefacts are pure finite volume effects and are proven to vanish in the $L\to \infty$ limit~\cite{Lucini:2015hfa}.
So, in this work, only the $C_3$ parts of the baryon correlation functions are computed. The masses that have been extracted from our smeared baryon interpolating operators are displayed in figure \ref{fig:baryons}.
\begin{figure}
    \centering
    \includegraphics[width=\textwidth]{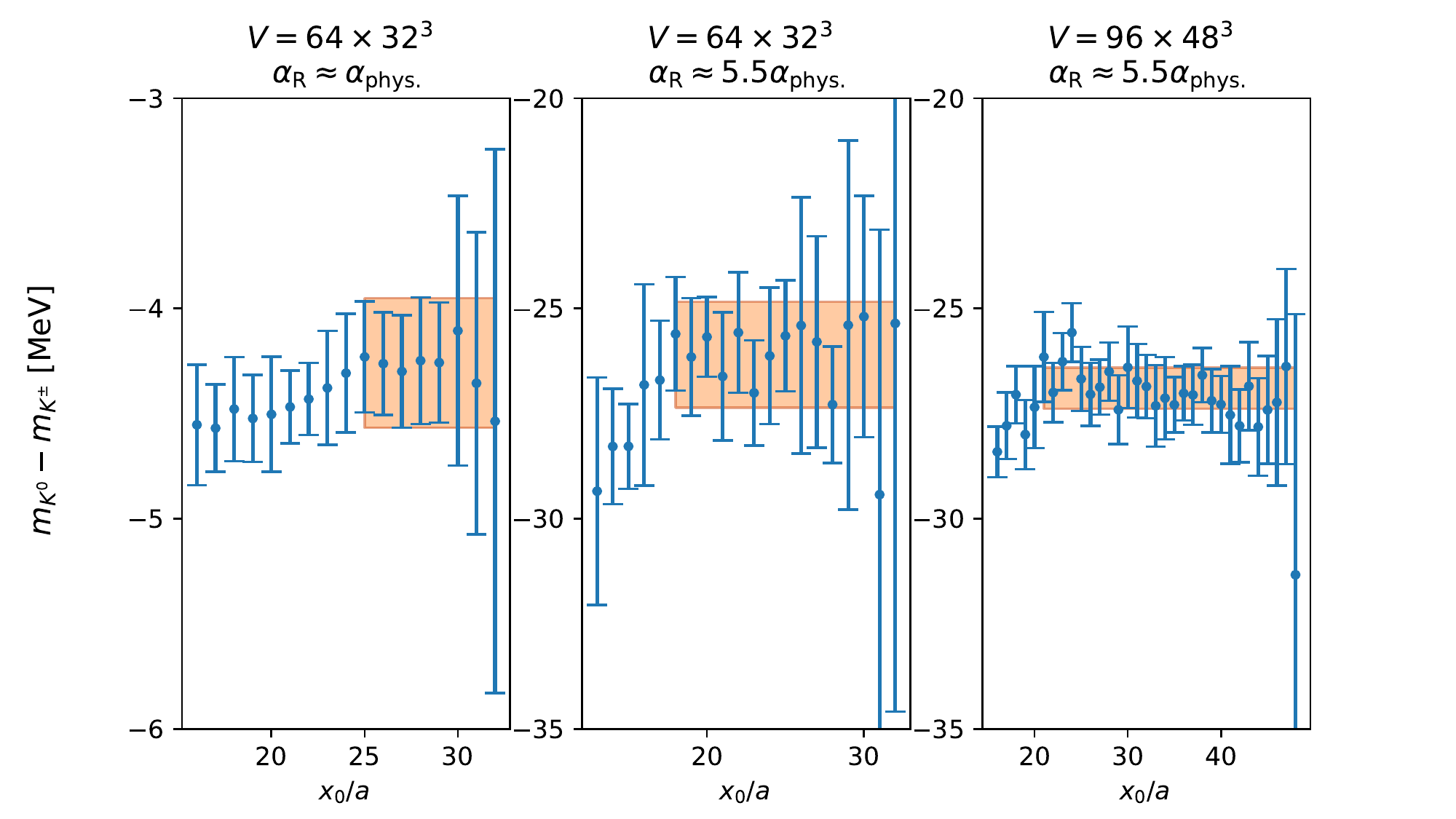}
    \caption{Splitting of the light mesons. In the left plot, one can see the splitting of the kaons on the ensemble A380. In the middle, the absolute value of the splitting is much larger as it was extracted from the ensemble A360. Keeping $\phi_2$ constant it follows that the splitting is proportional to $\alpha_\mathrm{R}$. Finally, on the right-hand side, the splitting is extracted on C380. The error bar is significantly smaller, compared to the smaller volume.
    In all plots, the orange error band denotes the effective mass and indicates the fit range that was used.}
    \label{fig:splitting}
\end{figure}
\begin{figure}
    \centering
    \begin{subfigure}[b]{0.5\textwidth}
        \centering
        \includegraphics[width=\textwidth]{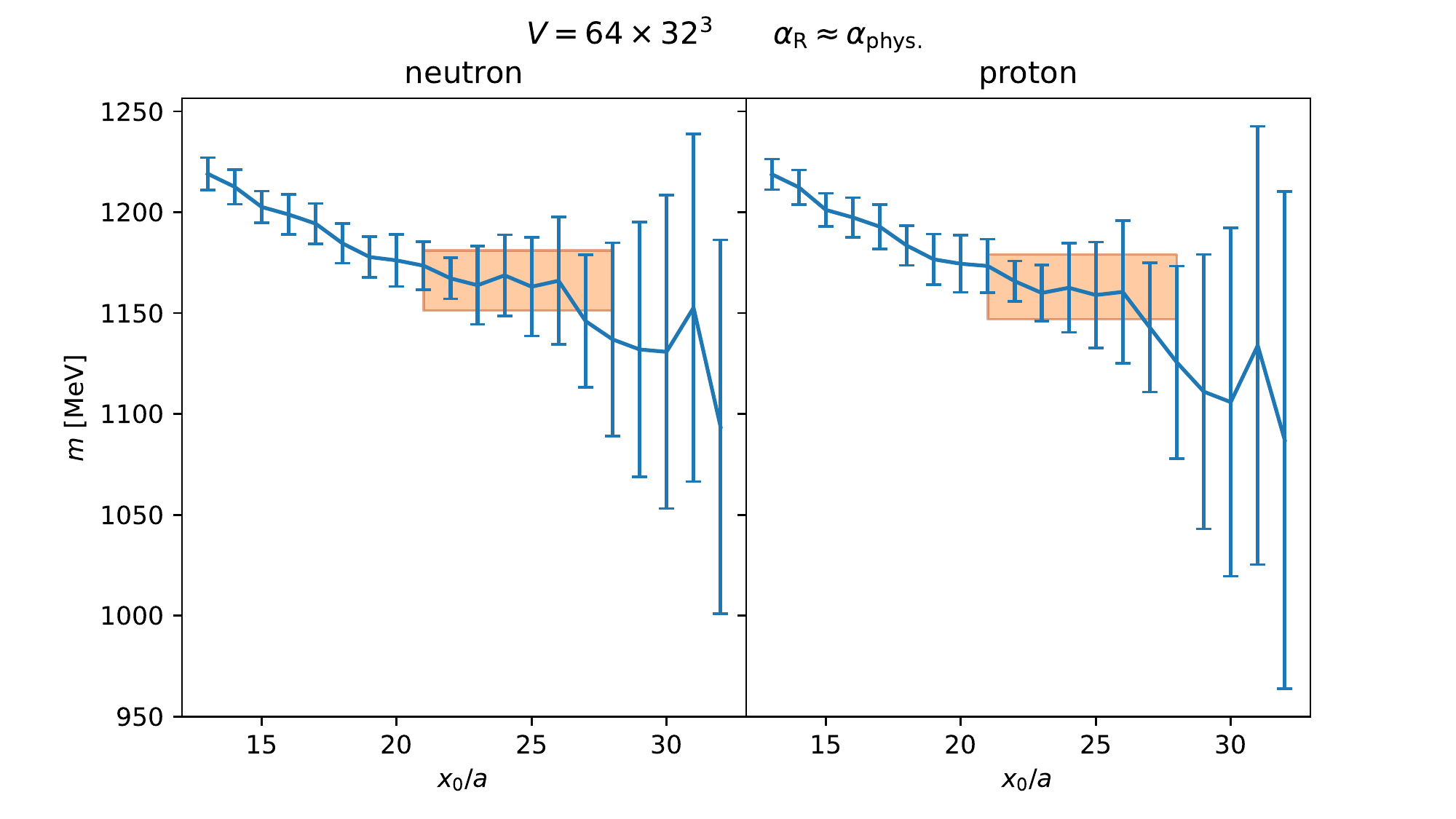}
        \caption{Proton and neutron effective masses.}
        \label{fig:proton_neutron}
    \end{subfigure}%
    \hfill%
    \begin{subfigure}[b]{0.5\textwidth}
        \centering
        \includegraphics[width=\textwidth]{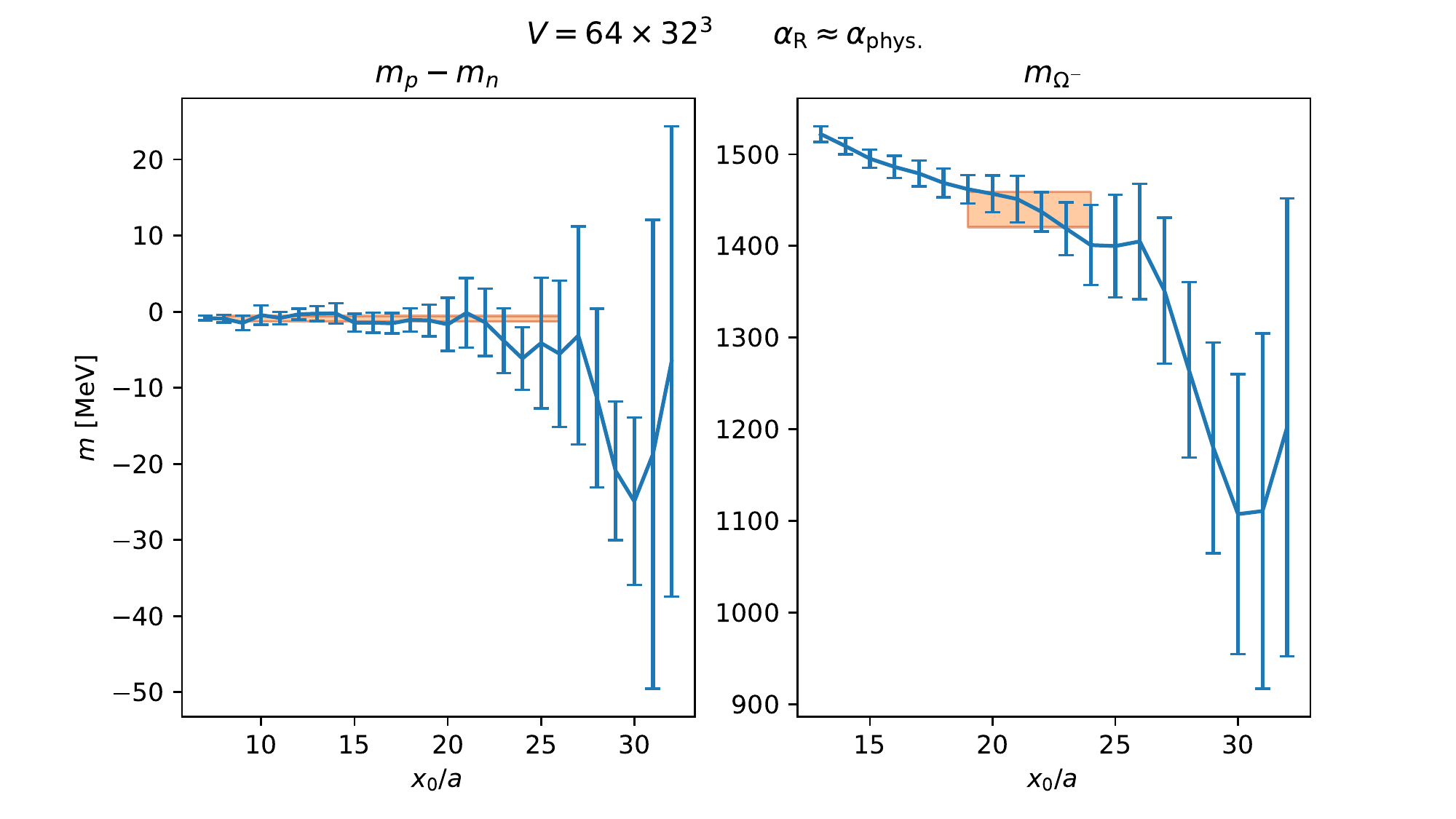}
        \caption{Proton and neutron splitting and omega effective mass.}
        \label{fig:omega}
    \end{subfigure}
    \caption{Baryon observables. Compared to the mesons in figure \ref{fig:splitting}, the baryon observables need to be extracted from intermediate times, since the large times are clearly dominated by noise. The small volume at physical electromagnetic coupling is currently not suitable to extract a signal for the nucleon splitting.}
    \label{fig:baryons}
\end{figure}

%
\section{Cost analysis}
\label{sec:cost}
At past conferences, the question about the cost of simulations with C$^\star$ boundary conditions was asked a couple of times. While we can not yet make an estimate of how the cost will scale to the physical point, we can compare a simulation of QCD with periodic boundary conditions with one with C$^\star$ boundaries. On top of that, we can give a number on the overhead of including dynamical QED in the simulations for the ensembles we have generated so far. In order to be able to compare the production cost, we have measured the time needed to generate a thermalized configuration on Lise\footnote{%
Lise has 1236 standard nodes with 384 GB memory, each of them with 2 CPUs. The
CPUs are Intel Cascade Lake Platinum 9242 (CLX-AP) with 48 cores each. The nodes
are connected with an Omni-Path network with a fat tree topology, 14 TB/s
bisection bandwidth, and 1.65 $\mu$s maximum latency. Source:
\url{https://www.hlrn.de/supercomputer-e/hlrn-iv-system/?lang=en}.
}
at HLRN for all gauge ensembles. The results are shown in table~\ref{tab:cost};
in particular, we report the specific cost, i.e. the cost in core$\times$seconds
per molecular dynamics unit divided by the number of lattice points.\footnote{%
The reader familiar with \texttt{openQ*D} knows that C$^\star$ boundary
conditions are implemented by means of an orbifold procedure which effectively
doubles the lattice size. In the code, one distinguishes between physical and
extended (i.e. doubled) lattice. Throughout this paper, we always refer to the
physical lattice. When we talk about lattice volume, we always refer to the
volume of the physical lattice. In particular, in order to reconstruct the total
cost from table~\ref{tab:cost}, one needs to multiply the specific cost times
the number of points of the physical lattice.
}
Comparing \texttt{A1} with \texttt{A400a00b324} one would naively expect that the latter was a factor of 2 more expensive since it uses C$^\star$- instead of periodic boundary conditions. Looking at the specific cost shows that the factor is in fact much smaller. This can be explained by a change in the algorithmic parameters. While both simulations used a three-level integrator and a similar number of fermionic forces, our ensemble \texttt{A400a00b324} integrates six instead of eight forces on the mid-level. This results in a factor of $6/8$ and reduces the cost gap. This gap is even further reduced by the fact that in the generation of \texttt{A1} smaller residues were used for the solvers. Comparing the number of times the Dirac operator is applied by the various solvers, we estimate the difference in residues to account for another factor of $0.84$. Thus, in total we estimate the specific cost of the QCD simulation with C$^\star$ boundary conditions to be $2 \times 6/8 \times 0.84 = 1.26$ times larger than the one with periodic boundary conditions. Comparing that with the actual factor of $0.42/0.35 = 1.2$ we can conclude that we have a good understanding of where the overhead comes from. Similar estimations for the other ensembles can be found in \cite{Bushnaq:2022aam}.
\begin{table}
   \small
   \begin{center}
   \begin{tabular}{cccc}
   \hline
   ensemble & global volume & n. cores
   & specific cost
   \\
   \rule{0mm}{4mm}
   & & 
   & $ \left[ \frac{\text{cores}\times\text{secs}}{\text{MDUs}\times\text{points}} \right]$
   \\[2mm]
   \hline
   \hline
   \texttt{A1} \cite{Hollwieser:2020qri} & $96 \times 32^3$ & 6144
   & 0.35 \\
   \hline
   \hline
   \texttt{A400a00b324} & $64 \times 32^3$ & 4096
   & 0.42\\
   \texttt{B400a00b324} & $80 \times 48^3$ & 2560
   & 0.62 \\
   \hline
   \texttt{A450a07b324} & $64 \times 32^3$ & 4096
   & 1.07 \\
   \texttt{A380a07b324} & $64 \times 32^3$ & 4096
   & 1.03 \\
   \hline
   \texttt{A500a50b324} & $64 \times 32^3$ & 4096
   & 0.88 \\
   \texttt{A360a50b324} & $64 \times 32^3$ & 4096
   & 1.05 \\
   \texttt{C380a50b324} & $96 \times 48^3$ & 3072
   & 1.40 \\
   \hline
   \end{tabular}
   \end{center}
   \caption{%
   Cost comparison of all production runs presented in this work with a
   $N_f=3+1$ QCD ensemble \texttt{A1} produced by the ALPHA
   collaboration~\cite{Hollwieser:2020qri}. For comparability, all times were measured on Lise at HLRN.
   }\label{tab:cost}
\end{table}

%
\section{Summary and outlook}
Up to the time of publication, we have generated five tuned QCD+QED ensembles at two different volumes. 
Comparing the cost of the generation of our current ensembles with QCD simulations at the SU(3)-symmetric point and periodic boundary conditions, we conclude that we understand the origin of the overhead of our simulations. We are currently generating another ensemble at $\alpha=0.02$. Unfortunately, we have also shown that the small-volume ensembles that we have generated suffer from considerable QED and QCD finite-volume effects. Hence in order to achieve an actual precise tuning, future simulations need to be done at much larger volumes. This is especially true once one wants to attempt to reach the physical point. In that case, the quarks, and thus the mesons used for tuning, will become much lighter and suffer from larger finite-volume effects. Switching to a physical scheme makes it also necessary to have a much more precise estimation of the omega mass. Hence, we want to improve the computation of the baryon masses by applying noise reduction techniques. 

Since the tuning and the generation of fully dynamical QCD+QED ensembles is rather expensive our collaboration is also investigating the RM123 method~\cite{deDivitiis:2013xla} in conjunction with C$^\star$ boundary conditions. Here, QED effects are included perturbatively in iso-symmetric QCD ensembles. This has the advantage, that the tuning is much simpler due to the smaller number of parameters. The disadvantage is that every new order in $\alpha$ poses a new challenge in the sense that the operator insertions in the correlations functions become more and more complicated. Investigating, how the fully dynamical approach and the RM123 method compare and what the limitations of each approach are is important not only from a conceptual but also from an economical point of view.


\paragraph{Acknowledgements.}
We gratefully acknowledge R. Höllwieser, F. Knechtli and T. Korzec for providing us with the information on the computational cost of their ensemble A1. Alessandro Cotellucci's and Jens L\"ucke's research is funded by the Deutsche Forschungsgemeinschaft (DFG, German Research Foundation) - Projektnummer 417533893/ GRK-2575 ``Rethinking Quantum Field Theory''. 
The funding from the European Union’s Horizon 2020 research and innovation program under grant agreements No. 813942 and No. 765048, as well as the financial support by SNSF (Project No. 200021\_200866) is grate- fully acknowledged. 
The authors gratefully acknowledge the computing time granted by the Resource
Allocation Board and provided on the supercomputer Lise and Emmy at NHR@ZIB and
NHR@Göttingen as part of the NHR infrastructure. The calculations for this
research were partly conducted with computing resources under the project
bep00085 and bep00102. The work was supported by CINECA that granted computing
resources on the Marconi supercomputer to the LQCD123 INFN theoretical
initiative under the CINECA-INFN agreement. The authors acknowledge access to
Piz Daint at the Swiss National Supercomputing Centre, Switzerland under the
ETHZ's share with the project IDs go22, go24, eth8, and s1101. The work was
supported by the Poznan Supercomputing and Networking Center (PSNC) through
grant numbers 450 and 466.

\bibliography{main.bib}

\begin{thebibliography}{18}
\expandafter\ifx\csname natexlab\endcsname\relax\def\natexlab#1{#1}\fi
\expandafter\ifx\csname url\endcsname\relax
  \def\url#1{{\tt #1}}\fi

\bibitem[Cotellucci et~al.(2023)]{RC:2022fly}
Alessandro Cotellucci et~al.
\newblock {Tuning of QCD+QED simulations with C$^\star$ boundary conditions}.
\newblock {\em PoS}, LATTICE2022:\penalty0 259, 2023, 2212.10894.

\bibitem[Kronfeld and Wiese(1991)]{Kronfeld:1990qu}
Andreas~S. Kronfeld and U.~J. Wiese.
\newblock {SU(N) gauge theories with C periodic boundary conditions. 1.
  Topological structure}.
\newblock {\em Nucl. Phys. B}, 357:\penalty0 521--533, 1991.

\bibitem[Kronfeld and Wiese(1993)]{Kronfeld:1992ae}
Andreas~S. Kronfeld and U.~J. Wiese.
\newblock {SU(N) gauge theories with C periodic boundary conditions. 2. Small
  volume dynamics}.
\newblock {\em Nucl. Phys. B}, 401:\penalty0 190--205, 1993, hep-lat/9210008.

\bibitem[Wiese(1992)]{Wiese:1991ku}
U.~J. Wiese.
\newblock {C periodic and G periodic QCD at finite temperature}.
\newblock {\em Nucl. Phys. B}, 375:\penalty0 45--66, 1992.

\bibitem[Polley(1993)]{Polley:1993bn}
L.~Polley.
\newblock {Boundaries for SU(3)(C) x U(1)-el lattice gauge theory with a
  chemical potential}.
\newblock {\em Z. Phys. C}, 59:\penalty0 105--108, 1993.

\bibitem[Lucini et~al.(2016)Lucini, Patella, Ramos, and
  Tantalo]{Lucini:2015hfa}
Biagio Lucini, Agostino Patella, Alberto Ramos, and Nazario Tantalo.
\newblock {Charged hadrons in local finite-volume QED+QCD with C$^{\star}$
  boundary conditions}.
\newblock {\em JHEP}, 02:\penalty0 076, 2016, 1509.01636.

\bibitem[Patella(2017)]{Patella:2017fgk}
Agostino Patella.
\newblock {QED Corrections to Hadronic Observables}.
\newblock {\em PoS}, LATTICE2016:\penalty0 020, 2017, 1702.03857.

\bibitem[Hansen et~al.(2018)Hansen, Lucini, Patella, and
  Tantalo]{Hansen:2018zre}
Martin Hansen, Biagio Lucini, Agostino Patella, and Nazario Tantalo.
\newblock {Gauge invariant determination of charged hadron masses}.
\newblock {\em JHEP}, 05:\penalty0 146, 2018, 1802.05474.

\bibitem[Campos et~al.(2018)Campos, Fritzsch, Hansen, Marinković, Patella,
  Ramos, and Tantalo]{openQxD-csic}
Isabel Campos, Patrick Fritzsch, Martin Hansen, Marina~Krstić Marinković,
  Agostino Patella, Alberto Ramos, and Nazario Tantalo.
\newblock openq*d, 2018.
\newblock URL \url{https://gitlab.com/rcstar/openQxD}.

\bibitem[Campos et~al.(2020)Campos, Fritzsch, Hansen, Marinkovic, Patella,
  Ramos, and Tantalo]{Campos:2019kgw}
Isabel Campos, Patrick Fritzsch, Martin Hansen, Marina~Krstic Marinkovic,
  Agostino Patella, Alberto Ramos, and Nazario Tantalo.
\newblock {openQ*D code: a versatile tool for QCD+QED simulations}.
\newblock {\em Eur. Phys. J. C}, 80\penalty0 (3):\penalty0 195, 2020,
  1908.11673.

\bibitem[L\"uscher(2016)]{openQCD}
Martin L\"uscher.
\newblock openqcd, 2016.
\newblock URL \url{https://cern.ch/luscher/openQCD}.

\bibitem[Altherr et~al.(2022)Altherr, Bushnaq, Campos, Catillo, Cotellucci,
  Dale, Fritzsch, Gruber, Luecke, Marinkovi\'c, Patella, Tantalo, and
  Tavella]{AlessandroPOS}
Anian Altherr, Lucius Bushnaq, Isabel Campos, Marco Catillo, Alessandro
  Cotellucci, Madeleine Evie~Beth Dale, Patrick Fritzsch, Roman Gruber, Jens
  Luecke, Marina~Krsti\'c Marinkovi\'c, Agostino Patella, Nazario Tantalo, and
  Paola Tavella.
\newblock {Tuning of QCD+QED simulations with C$^{\star}$ boundary conditions}.
\newblock {\em To appear in PoS}, LATTICE2022, 2022.

\bibitem[Luecke et~al.(2022)Luecke, Bushnaq, Campos, Catillo, Cotellucci, Dale,
  Fritzsch, Marinkovi\'c, Patella, and Tantalo]{RC:2021tah}
Jens Luecke, Lucius Bushnaq, Isabel Campos, Marco Catillo, Alessandro
  Cotellucci, Madeleine Evie~Beth Dale, Patrick Fritzsch, Marina~Krsti\'c
  Marinkovi\'c, Agostino Patella, and Nazario Tantalo.
\newblock {An update on QCD+QED simulations with C$^\star$ boundary
  conditions}.
\newblock {\em PoS}, LATTICE2021:\penalty0 293, 2022, 2108.11989.

\bibitem[Bushnaq et~al.(2022)Bushnaq, Campos, Catillo, Cotellucci, Dale,
  Fritzsch, L\"ucke, Krsti\'c~Marinkovi\'c, Patella, and
  Tantalo]{Bushnaq:2022aam}
Lucius Bushnaq, Isabel Campos, Marco Catillo, Alessandro Cotellucci, Madeleine
  Dale, Patrick Fritzsch, Jens L\"ucke, Marina Krsti\'c~Marinkovi\'c, Agostino
  Patella, and Nazario Tantalo.
\newblock {First results on QCD+QED with C$^\star$ boundary conditions}.
\newblock 9 2022, 2209.13183.

\bibitem[Zyla et~al.(2020)]{ParticleDataGroup:2020ssz}
P.~A. Zyla et~al.
\newblock {Review of Particle Physics}.
\newblock {\em PTEP}, 2020\penalty0 (8):\penalty0 083C01, 2020.

\bibitem[Bruno et~al.(2017)Bruno, Korzec, and Schaefer]{Bruno:2016plf}
Mattia Bruno, Tomasz Korzec, and Stefan Schaefer.
\newblock {Setting the scale for the CLS $2 + 1$ flavor ensembles}.
\newblock {\em Phys. Rev. D}, 95\penalty0 (7):\penalty0 074504, 2017,
  1608.08900.

\bibitem[H\"ollwieser et~al.(2020)H\"ollwieser, Knechtli, and
  Korzec]{Hollwieser:2020qri}
Roman H\"ollwieser, Francesco Knechtli, and Tomasz Korzec.
\newblock {Scale setting for $N_f=3+1$ QCD}.
\newblock {\em Eur. Phys. J. C}, 80\penalty0 (4):\penalty0 349, 2020,
  2002.02866.

\bibitem[de~Divitiis et~al.(2013)de~Divitiis, Frezzotti, Lubicz, Martinelli,
  Petronzio, Rossi, Sanfilippo, Simula, and Tantalo]{deDivitiis:2013xla}
G.~M. de~Divitiis, R.~Frezzotti, V.~Lubicz, G.~Martinelli, R.~Petronzio, G.~C.
  Rossi, F.~Sanfilippo, S.~Simula, and N.~Tantalo.
\newblock {Leading isospin breaking effects on the lattice}.
\newblock {\em Phys. Rev. D}, 87\penalty0 (11):\penalty0 114505, 2013,
  1303.4896.

\end{thebibliography}

\end{document}